\documentclass[aps,pra,reprint,groupedaddress,showpacs]{revtex4-1}

\usepackage{graphicx}
\usepackage{amssymb,amsmath}
\usepackage{bm}
\usepackage{bbold} 
\usepackage{multirow}

\begin{document}

\title{Proton-hydrogen collision at cold temperatures}

\author{Ming Li}
\email[]{ming.li3@rockets.utoledo.edu}
\author{Bo Gao}
\email[]{bo.gao@utoledo.edu}
\affiliation{Department of Physics and Astronomy,
    University of Toledo, Mailstop 111,
    Toledo, Ohio 43606,
    USA}

\date{\today}

\begin{abstract}

We study the proton-hydrogen collision in the energy range from 0 to 5 K where the hyperfine structure of the hydrogen atom is important. The proper multichannel treatment of the hyperfine structure is found to be crucial at cold temperatures compared to the elastic approximation used for higher temperatures. Both elastic and hyperfine-changing inelastic processes are investigated using the newly developed multichannel quantum-defect theory (MQDT) as well as the traditional coupled-channel numerical method. Cross section results from the two methods coincide well throughout the energy range. The MQDT with a few parameters provide an efficient and basically analytic description of the proton-hydrogen interaction in this energy regime.



\end{abstract}


\maketitle

\section{Introduction}
\label{sec:Intro}

Scattering properties of the proton-hydrogen interaction have been investigated theoretically throughout the last six decades with increasing accuracy and expanding energy range~\cite{Dal53, Bat62, Smith1966, Hunter77, Hod91, Krstic2004, bod08}. Such data are important for the understanding of the physics in planetary atmospheres~\cite{Hod91, Huestis2008} and in intergalactic media~\cite{Mei09}, especially for the interpretation of the brightness of the $21$cm transition, which depends essentially on the spin temperature of atomic hydrogens~\cite{Glassgold05, Furlanetto2007, Huestis2008, Mei09}. 

While the past studies have covered most of the energy range that is of astrophysical interest, one uncharted territory is the regime of cold and ultracold temperatures. The lowest energy studied by previous works is $10^{-4}$ eV~\cite{Hunter77, Krstic2004} (equivalent to $1.16$ K), which is about an order of magnitude larger than the hyperfine splitting of the atomic hydrogen (equivalent to about $0.07$K), using the elastic approximation~\cite{dal61} that neglects the hyperfine interaction. To go beyond this energy into the cold and ultracold temperature regime, the hyperfine interaction is expected to play an important role while the elastic approximation breaks down. This has been shown for the hydrogen-hydrogen scattering problem using multichannel numerical method which rigorously incorporated the hyperfine interaction~\cite{zyg03}. To address this issue to produce new accurate results for cold and ultracold temperatures, we use the multichannel quantum defect theory (MQDT) for ion-atom interactions developed in our previous work~\cite{LYG2014}, which provides an almost analytic description of the interaction with only a handful of parameters throughout the whole energy range. Since MQDT is a newly developed theory, we also present results using the more traditional multichannel numerical method as a comparison and reassurance.

In Sec.~\ref{sec:Theory}, we go over the multichannel theory for the proton-hydrogen interaction with MQDT. The multichannel numerical method and the elastic approximation are also briefly illustrated, which are used for comparisons. In Sec.~\ref{sec:MvE}, we show the comparisons of cross section results from the multichannel treatment and the elastic approximation from $30$ kelvin down to $0.1$ kelvin, within which temperature range the elastic approximation starts to fail. Then in Sec.~\ref{sec:NvM}, we present the new cross section results in the cold and ultracold regime using MQDT, with results also from the multichannel numerical calculation as comparisons.

\section{Theory}
\label{sec:Theory}

The scattering process of a proton and a hydrogen atom in its ground $^2S$ electronic state falls into the category of the ion-atom interaction of the type $^1S+^2S$ with identical nuclei of spin $I_2=I_1=1/2$, which was studied in Ref.~\cite{LYG2014}. With consideration of the hyperfine interaction, the process can be expressed as
\begin{align}
H(F_{1i},M_{1i})&+H^+(F_{2},M_{2i}) \nonumber \\
&\longrightarrow H(F_{1j},M_{1j})+H^+(F_{2},M_{2j}) \;.
\label{eq:reaction}
\end{align}
Here $F_1=I_1 \pm 1/2$ is the asymptotic total angular momentum corresponding to the $^2S$ electronic state of the hydrogen atom, and $F_2=I_2=I_1$ is the asymptotic total angular momentum of the proton. The $M$s are the corresponding magnetic quantum numbers. The sub-indices $i$ and $j$ refer to the internal states before and after the collision respectively. Notice that this process can alter the magnitude of the asymptotic total angular momentum of the hydrogen atom, as well as both the magnetic quantum numbers of the hydrogen atom and the proton. The asymptotic total quantum angular momentum of the proton $F_2$ does not change. Here we assume that the nuclear spins do not change in the collision, because their coupling with other angular momenta is weak compared to other couplings when the two nuclei are close, and is only important when the two nuclei are well seperated which gives rise to the atomic hyperfine splitting. Thus the change in the atomic hyperfine level is mainly due to the ``exchange'' of the electron between the two nuclei. The following theory can be easily applied to other ion-atom interactions with $^2S+^1S$ type of electronic states and identical nuclei~\cite{LYG2014}.

The Hamiltonian describing the system of interest can be written as
\begin{equation}
H = -\frac{\hbar^2}{2\mu}\triangledown_R^2 + H_{\textrm{BO}} + H_{\textrm{hf}} \;,
\label{eq:hamiltonian}
\end{equation}
where $\mu$ is the reduced mass and $R$ is the internuclear distance. $H_{\textrm{hf}}$ represents the hyperfine interaction and $H_{\textrm{BO}}$ is the adiabatic Born-Oppenheimer (BO) Hamiltonian. In our case of a slow collision that only involves the $S$ electron, there are two relevant BO potential energy curves characterized by $^2\Sigma_{g,u}^+$ (or $1s\sigma_g$ and $2p\sigma_u$ states).

How to treat the hyperfine term effectively and efficiently depends on the importance of the hyperfine interaction compared to the BO Hamiltonian and the kinetic energy term in Eq.~(\ref{eq:hamiltonian}). When the asymptotic kinetic energy is comparable or smaller than the hyperfine interaction energy, the hyperfine term has to be taken into consideration, especially for the purpose of obtaining the correct asymptotic thresholds, where it can be approximately replaced by the atomic hyperfine interaction as the two nuclei are well separated. When the two nuclei are close, the BO term dominates over the hyperfine interaction, and the hyperfine term can be neglected to simplify the theory.
The other case is when the asymptotic kinetic energy is much larger than the hyperfine interaction energy, and the elastic approximation can be applied~\cite{dal61, cot00, LG2012} by neglecting the hyperfine term regardless of the nuclear separation.

In subsection~\ref{sec:radial}, we show the decomposition of the wavefunction and the derivation of the radial coupled-channel (CC) equations. Subsection~\ref{sec:channel} illustrates the channel structure of the system with two different angular momentum coupling schemes. The frame transformation between the two sets of coupling basis functions is also given in this subsection. In subsection~\ref{sec:cs}, we present the scattering amplitude in terms of the $S$ matrix with the proper symmetry consideration for identical nuclei, and various cross sections that can be derived from it. To obtain the $S$ matrix from the CC equations, we present two methods here: MQDT and the more traditional numerical method. In subsection~\ref{sec:mqdt}, MQDT formulation is introduced and explained in detail, as well as the determination of the QDT parameters. Subsection~\ref{sec:numerical} shows the procedure to construct the potential energy curves for numerical calculation and the numerical method used in this work. In subsection~\ref{sec:elastic}, we briefly go over the elastic approximation and present the simplified expressions of the total cross sections under the elastic approximation to be examined later.

\subsection{Radial coupled-channel equations}
\label{sec:radial}

The total angular momentum $\bm{T}$ in the proton-hydrogen interaction is conserved. It can be split into two parts as $\bm{T} = \bm{F} + \bm{l}$, where $\bm{F}$ is the total ``spin'' angular momentum and $\bm{l}$ is the relative orbital angular momentum of the nuclei. In interactions that involve only $S$ electrons, $\bm{F}$ and $\bm{l}$ are independently conserved, in which case, the wave function, for each $\bm{F}$ and $\bm{l}$ can be written as 
\begin{equation}
\psi^{FMlm} = \left[\sum_{b}\Phi^{FM}_b(\bm{R}) G^{Fl}_b(R)/R\right] Y_{lm}(\hat{\bm{R}}) \;.
\label{eq:wavefn}
\end{equation}
Here $M$ and $m$ are projections of $\bm{F}$ and $\bm{l}$ on a space-fixed axis, respectively. $\Phi^{FM}_b(\bm{R})$ is the channel function of channel $b$ that belongs to an $F$ manifold (the sub-index $a$, used later, also represents a channel that comes from the same set of channels as channel $b$), $G^{Fl}_b(R)$ is the corresponding radial wave function, and $Y_{lm}(\hat{\bm{R}})$ is the spherical harmonics. The angular momentum coupling schemes that define different sets of channel functions are described in the next subsection.

We substitute Eq.~(\ref{eq:wavefn}) into the Schr\"{o}dinger equation,
\begin{equation}
H\psi^{FMlm} = \epsilon\psi^{FMlm} \;.
\label{eq:schrodinger}
\end{equation}
Upon using the orthogonality properties of the channel functions and ignoring nonadiabatic couplings, we arrive at a set of CC equations
\begin{align}
\left( -\frac{\hbar^2}{2\mu}\frac{\mathrm{d}^2}{\mathrm{d}R^2}\right.& \left.+\;\frac{l(l+1)\hbar^2}{2\mu R^2}-\epsilon \right) G^{Fl}_a(R) \nonumber \\
& +\sum_b [V^{\textrm{BO}}_{ab}(R) + V^{\textrm{hf}}_{ab}(R)]G^{Fl}_b(R) = 0 \;,
\label{eq:CC}
\end{align}
where
\begin{subequations}
\begin{align}
V^{\textrm{BO}}_{ab}(R) &\equiv \langle \Phi^{FM}_a|H_{\textrm{BO}}|\Phi^{FM}_b\rangle \;,
\label{eq:Vdef1}\\
V^{\textrm{hf}}_{ab}(R) &\equiv \langle \Phi^{FM}_a|H_{\textrm{hf}}|\Phi^{FM}_b\rangle \;,
\label{eq:Vdef2}
\end{align}
\label{eq:Vdef}
\end{subequations}
which are independent of $M$. 

\subsection{Channel structure and frame transformation}
\label{sec:channel}

\begin{table}
\caption{Channel structure for the proton-hydrogen interaction in the ground electronic states. 
    \label{tb:channels}}
\begin{ruledtabular}
\begin{tabular}{ccc}
\multirow{2}{*}{Total $F$} & $FF$ coupling & $JI$ coupling \\ & $\{F_1,F_2\}$ & $\{J,I\}$ \\
\hline
\multirow{2}{*}{$1/2\le F \le 2I_1-1/2$} & $\{I_1-1/2,I_1\}$ & $\{1/2,F-1/2\}$ \\
 & $\{I_1+1/2,I_1\}$ & $\{1/2,F+1/2\}$ \\
\hline $F=2I_1+1/2$ & $\{I_1+1/2,I_1\}$ & $\{1/2,2I_1\}$
\end{tabular}
\end{ruledtabular}
\end{table}

For each $l$, the total number of states is $2(2I_1+1)^2 = 8$, and the total number of channels is $4I_1+1 = 3$. These three channels are separated into two uncoupled groups for the two different $F$s. For $1/2\le F \le 2I_1-1/2$, which makes $F = 1/2$, there are two coupled channels. For $F=2I_1+1/2=3/2$, there is only one channel and only elastic scattering can happen.

We adopt two angular momentum coupling schemes to decompose $\bm{F}$, following the theory of Ref.~\cite{gao96}. The $FF$ coupling scheme, $\bm{F}=\bm{F}_1+\bm{F}_2$, forms the fragmentation channels, or $FF$ channels, that diagonalize the total Hamiltonian when $R$ goes to infinity. The scattering boundary conditions, hence the $S$ matrix, are defined in the $FF$ channels. The $JI$ coupling scheme, $\bm{F}=\bm{J}+\bm{I}$, forms the condensation channels, or $JI$ channels, that diagonalize the adiabatic BO Hamiltonian, hence approximately diagonalize the total Hamiltonian in the short range due to the insignificance of the hyperfine interaction in that region. $\bm{J}$ is the total electronic angular momentum, which can only have the magnitude of $1/2$ here, and $\bm{I}$ is the total nuclear spin. The $JI$ channels are most directly related to the BO potential curves $^2\Sigma_{g,u}^+$, and are channels in which the short-range $K^c$ matrix of the MQDT formulation \cite{LYG2014} has the simplest representation. 
The detailed channel structure is illustrated in Table~\ref{tb:channels}.

For $F = 1/2$, the asymptotic thresholds $E_1$ and $E_2$ of the two coupled channels in the $FF$ coupling basis are separated by the atomic hyperfine splitting $E_2 - E_1 = \Delta E^{\textrm{hf}}$, known as the $21$ cm line for hydrogen atoms. It is given by $\Delta E^{\textrm{hf}}/h \approx 1420.405751768$ MHz \cite{Hel70} ($\Delta E^{\textrm{hf}}/k_B \approx 0.068168729$ K). Following the theory of Ref.~\cite{gao96}, the $FF$ and the $JI$ coupling basis functions are related by a frame transformation given by a two by two orthogonal matrix
\begin{widetext}
\begin{equation}
U^{F} = \frac{(-1)^{2F+1}}{\sqrt{2(2I_1+1)}} 
	\left(
	\begin{array}{cc}
	-\sqrt{2I_1-F+1/2} & \sqrt{2I_1+F+3/2}\\
	\sqrt{2I_1+F+3/2} & \sqrt{2I_1-F+1/2}
	\end{array}
	\right) \;.
\label{eq:frame}
\end{equation}
\end{widetext}
The ordering of the channels here as well as in the matrices later in this paper follows the ordering in Table~\ref{tb:channels}.


\subsection{Scattering amplitude and cross sections}
\label{sec:cs}

The scattering amplitude for processes of Eq.~(\ref{eq:reaction}) that satisfies the scattering boundary conditions in the $FF$ channels are given by \cite{gao96}
\begin{align}
 &f\left(\{F_{1i}M_{1i}, F_2M_{2i}\} \bm{k}_i
	\rightarrow \{F_{1j}M_{1j}, F_2M_{2j}\} \bm{k}_j \right) \nonumber\\
= &-\sum_{lmFM} \frac{2\pi i}{(k_ik_j)^{1/2}}
	Y^*_{lm}(\hat{k}_i)Y_{lm}(\hat{k}_j) \nonumber\\
	&\times \langle F_{1j}M_{1j},F_{2}M_{2j}|FM_F\rangle
	\left[S^{Fl}(\epsilon)-\mathbb{1}\right]_{ji} \nonumber\\
	&\times\langle FM_F|F_{1i}M_{1i},F_{2}M_{2i}\rangle \;,
\label{eq:scamp}
\end{align}
where $\mathbb{1}$ is the unit matrix and $S^{Fl}$ is the $S$ matrix defined in the $FF$ channels~\cite{gao96}. $\hbar\bm{k}_{i, j}$ are the initial and the final relative momenta in the center-of-mass frame. 

Notice that the subscript labelling of $1$ or $2$ in the scattering amplitude equation refers to the individual quantum states, but not the nuclei, because the nuclei are identical. If we use $f(i\rightarrow j, \bm{k}_j)$ as a short-hand notation for the amplitude of Eq.~(\ref{eq:scamp}), the differential cross section of the process of $i\rightarrow j$ at angle $\hat{\bm{k}}_j$ has both contributions from $f(i\rightarrow j, \bm{k}_j)$, for particle with state $(F_{2}, M_{2j})$ to be detected, and $f(i\rightarrow j, -\bm{k}_j)$, for particle with state  $(F_{1j}, M_{1j})$ to be detected. Thus, the differential cross section that carefully takes account of the symmetry property of identical nuclei is given by \cite{gao96,Gao2013b}
\begin{multline}
\frac{d\sigma}{d\Omega_j}
	\left(\{F_{1i}M_{1i}, F_2M_{2i}\} \bm{k}_i
	\rightarrow \{F_{1j}M_{1j}, F_2M_{2j}\} \bm{k}_j \right) \\
=\frac{k_j}{k_i}\frac{1}{2}\left(\left|f(i\rightarrow j,\bm{k}_j)\right|^2
	+\left|f(i\rightarrow j,-\bm{k}_j)\right|^2\right)\;.
\label{eq:dfxs}
\end{multline}
Notice that the differential cross section is same for direction $\hat{\bm{k}}_j$ and $-\hat{\bm{k}}_j$.

Many different cross sections can be derived from the differential cross section. In particular, the total cross sections for elastic (include $M$-changing) collisions, and the hyperfine excitation and de-excitation processes, after averaging over initial states and summing over final states, are given by
\begin{align}
\sigma(\{F_{1i},F_{2}\}&\rightarrow\{F_{1j},F_{2}\}) = \frac{\pi}{(2F_{1i}+1)(2F_{2}+1)k_i^2}  \nonumber\\
&\times \sum_{Fl}(2l+1)(2F+1) |S^{Fl}_{ji}-\delta_{ji}|^2 \;.
\label{eq:totalcs}
\end{align}
The next step is to solve the CC equations to obtain the $S$ matrix to be used in these cross section equations.

\subsection{MQDT}
\label{sec:mqdt}

One way to obtain the $S$ matrix is MQDT. The MQDT for ion-atom interactions, as demonstrated in Ref.~\cite{LYG2014}, consists of the formulation of Ref.~\cite{gao05a} in combination with the QDT functions for the $-1/R^4$-type potentials as detailed in Ref.~\cite{Gao2013c}. It takes full advantage of the physics that both the energy dependence \cite{gao98b} and the partial wave dependence \cite{gao01} of the atomic interaction around a threshold are dominated by effects of the long-range potential, which are encapsulated in the universal QDT functions. The short-range contribution is isolated to a short-range $K^c$ matrix that is insensitive to both the energy and the partial wave. 

\subsubsection{General formulation}

For an $N$-channel problem at energies where all channels are open, the MQDT gives the physical $K$ matrix, in our case the $K^{Fl}$, as~\cite{gao05a}
\begin{equation}
K^{Fl} = -(Z^c_{fc}-Z^c_{gc}K^c)(Z^c_{fs} - Z^c_{gs}K^c)^{-1} \;,
\label{eq:Kphyo}
\end{equation}
where $Z^c_{xy}$s are $N\times N$ diagonal matrices with elements $Z^c_{xy}(\epsilon_{si},l)$ being the $Z^c_{xy}$ functions \cite{Gao2013c} evaluated at the scaled energy $\epsilon_{si}=(\epsilon-\epsilon_i)/s_E$ relative to the respective channel threshold $\epsilon_i$. Here $s_E = (\hbar^2/2\mu)(1/\beta_4)^2$ and $\beta_4 = (\mu \alpha_A/\hbar^2)^{1/2}$ are the characteristic energy and the length scales, respectively, associated with the polarization potential, $-\alpha_A/2R^4$, with $\alpha_A$ being the static polarizability of the atom. 

At energies where $N_o$ channels are open, and $N_c=N-N_o$ channels are closed, the MQDT gives \cite{gao05a}
\begin{equation} 
K^{Fl} = -(Z^c_{fc}-Z^c_{gc}K^c_{\mathrm{eff}})(Z^c_{fs} - Z^c_{gs}K^c_{\mathrm{eff}})^{-1} \;,
\label{eq:Kphy}
\end{equation}
where
\begin{equation}
K^c_{\mathrm{eff}} = K^c_{oo}+K^c_{oc}(\chi^c -K^c_{cc})^{-1}K^c_{co} \;,
\label{eq:Kceff}
\end{equation}
in which $\chi^c$ is a $N_c\times N_c$ diagonal matrix with elements $\chi^c_l(\epsilon_{si},l)$ \cite{Gao2013c}, and $K^{c}_{oo}$, $K^{c}_{oc}$, $K^{c}_{co}$, and $K^{c}_{cc}$, are submatrices of $K^{c}$ corresponding to open-open, open-closed, closed-open, and closed-closed channels, respectively. Equation~(\ref{eq:Kphy}) is formally the same as Eq.~(\ref{eq:Kphyo}), except that the $K^c$ matrix is replaced by the $K^c_{\mathrm{eff}}$ that accounts for the effects of the closed channels. 

From the physical $K$ matrix, the $S$ matrix is obtained as~\cite{gao96}
\begin{equation}
S^{Fl} = (\mathbb{1}+i K^{Fl})(\mathbb{1}-i K^{Fl})^{-1} \;.
\label{eq:Smatrix}
\end{equation} 

\subsubsection{$K^c$ matrix and the short-range parametrization}

The short-range $K^c$ matrix has only two independent elements, which are two slowly varying functions of energy and $l$, the single-channel $K^c$ matrices $K^c_{g, u}(\epsilon, l)$ that represent the $g$ and $u$ adiabatic BO molecular states. They are directly related to the corresponding quantum defects by~\cite{Gao2013c}
\begin{equation}
K^c_{g,u} (\epsilon, l) = \mathrm{tan}[\pi\mu^c_{g,u}(\epsilon, l)+\pi/4] \;.
\label{eq:muctoKc}
\end{equation} 

For $F = 1/2$ where there are two coupled channels, the $K^c$ matrix in the $FF$ channels can be obtained from the one in the $JI$ channels through the frame transformation, given by
\begin{equation}
K^c = U^{F\dagger}K^{c(JI)}U^{F} \;,
\end{equation}
where
\begin{equation}
K^{c(JI)} =
	\left(
	\begin{array}{cc}
	\frac{(K^c_g+K^c_u) + {e_2}(K^c_g-K^c_u)}{2}& 0 \\
	0 & \frac{(K^c_g+K^c_u) - {e_2}(K^c_g-K^c_u)}{2}
	\end{array}
	\right) \;,
\label{eq:KcJI}
\end{equation}
where $e_2=(-1)^{F+l-1/2}=(-1)^l$. For $F = 3/2$ where there is only one channel, $K^c$ is either given by $K^c_{g}$ or $K^c_{u}$ depending on whether $e_1=(-1)^{2I_1+l}=(-1)^{l+1}$ is positive or negative:
\begin{equation}
K^c = \frac{1}{2}\left[(K^c_g+K^c_u) + {e_1}(K^c_g-K^c_u)\right] \;.
\end{equation}

In the simplest MQDT implementation, instead of two full potential curves used in the numerical calculation, only three constant parameters are needed besides the hyperfine splitting and the reduced mass. The static dipole polarizability of the hydrogen atom $\alpha_A=9/2$ a.u.~\cite{Damburg1968} characterizes the long range part of the potential. The two QDT parameters, the zero energy zero angular momentum single-channel $K^c_{g,u}(0,0)$, characterize the short range part of the potential due to the energy and partial wave insensitive nature of the short range interaction. They are related to the corresponding $s$ wave scattering lengths by~\cite{gao03,gao04b}
\begin{equation}
a^{g,u}_{l=0}/\beta_n = \left(b^{2b}\frac{\Gamma(1-b)}{\Gamma(1+b)}\right)
    \frac{K^c_{g,u}(0,0) + \tan(\pi b/2)}{K^c_{g,u}(0,0) - \tan(\pi b/2)} \;,
\label{eq:a0sKc}
\end{equation}
where $b=1/(n-2)$. It reduces to, for $n=4$~\cite{LG2012}, 
\begin{equation}
a^{g,u}_{l=0}/\beta_4 = \frac{K^c_{g,u}(0,0) + 1}{K^c_{g,u}(0,0) - 1} \;.
\label{eq:a0sKc4}
\end{equation}

More accurate results over a greater range of energies can be obtained by incorporating the energy dependence, and especially, for the range of energy under consideration, the partial wave dependence of the short-range parameters~\cite{LYG2014}. These weak dependences are well described by expansions
\begin{equation}
\mu^c_{g, u} (\epsilon, l) \approx \mu^c_{g, u} (0, 0) + b^{\mu}_{g,u}
\epsilon+c^{\mu}_{g, u} [l(l + 1)]  \;,
\label{eq:mucel}
\end{equation}
in which the parameters $b^{\mu}_{g,u}$ and $c^{\mu}_{g, u}$ characterize the energy and the partial wave dependences of the quantum defects for the \textit{gerade} and \textit{ungerade} states, respectively. 

\subsubsection{Determination of QDT parameters} 


\begin{table}
\caption{Zero energy QDT parameters used for comparison with numerical calculations 
    \label{tb:QDTpara}}
\begin{ruledtabular}
\begin{tabular}{ccccc}
\textit{g} or \textit{u} & $K^c(0,0)$ & $\mu^c(0,0)$ & $a_{l=0}$ (a.u.) & $c^\mu$ \\
\hline
\textit{gerade} & -0.35829 & 0.64049 & -30.371 & 0.0116 \\
\textit{ungerade} & 1.1723 & 0.025194 & 810.52 & 0.0214 \\
\end{tabular}
\end{ruledtabular}
\end{table}

The simplest MQDT implementation works well for the first handful of partial waves for hydrogen but starts to show noticeable deviation from the numerical calculation when $l$ becomes large, thus it breaks down at energies where higher partial waves start to contribute significantly. To parametrize the short-range interaction more accurately for energies ranging from ultracold temperatures all the way to 5 kelvin, we need to use the expansion of the quantum defects from Eq.~(\ref{eq:mucel}). 
 
The zero energy zero partial wave short-range parameters as well as parameters $b^{\mu}_{g,u}$ and $c^{\mu}_{g, u}$ can be determined easily by solving the single-channel radial equations with $V_{g,u}$ as the potential terms at a few energies for a few partial waves. For the energy range considered here, the energy dependence of $\mu^c$ is found to be negligible, namely $b^{\mu}_{g,u}\approx 0$. To determine the rest of the parameters, we first, for the first several partial waves at zero energy, propagate the radial wavefunctions through single-channel calculations and match them to the proper scattering boundary conditions, which is given by
\begin{equation}
u_{\epsilon l}(R) = A_{\epsilon l}[f^c_{\epsilon l}(R)-K^c(\epsilon, l)g^c_{\epsilon l}(R)] \;, 
\label{eq:kcdef}
\end{equation}
at progressively larger $R$ until the resulting $K^c_{g,u}$ converge to a desired accuracy \cite{gao01,gao03,gao08a,LG2012}. Here $f^c$ and $g^c$ at $\epsilon = 0$ are the zero-energy QDT reference functions for the $-1/R^4$ potential given in Ref.~\cite{gao04b,gao08a}. Then the resulting $K^c_{g,u}(\epsilon=0,l)$ are converted into $\mu^c_{g, u}(\epsilon=0, l)$ and fit into Eq.~(\ref{eq:mucel}) for various partial waves. The parameters obtained using the potential energy curves constructed in the next subsection are listed in Table~\ref{tb:QDTpara}. The $s$-wave scattering lengths are calculated using Eq.~(\ref{eq:a0sKc4}) from the corresponding single-channel $K^c$s. They are in good agreement with the previous results~\cite{Glassgold05, bod08, Kai13}.

\subsection{Potential energy curves and numerical method}
\label{sec:numerical}

In order to numerically calculate the $S$ matrix to compare to the MQDT results, and also to extract the QDT parameters in this paper, we need to construct the potential energy terms in Eqs.~(\ref{eq:Vdef1}) and (\ref{eq:Vdef2}) in the $FF$ channels. 

For $F = 1/2$, there are two coupled channels. The hyperfine term can be approximated as diagonal and constant in the $FF$ channels, which is given by
\begin{equation}
V^{\textrm{hf}} = 
	\left(
	\begin{array}{cc}
	0 & 0 \\
	0 & \Delta E^{\textrm{hf}}
	\end{array}
	\right) \;.
\label{eq:Vhf}
\end{equation}
The BO potential energy matrix in the $FF$ channels is given in terms of the matrix in the $JI$ channels with a frame transformation, as
\begin{equation}
V^{\textrm{BO}} = U^{F\dagger}V^{\textrm{BO}(JI)}U^{F} \;.
\label{eq:VBOFF}
\end{equation}
The BO potential energy matrix in the $JI$ channels is diagonal and can be written as
\begin{equation}
V^{\textrm{BO}(JI)} =
	\left(
	\begin{array}{cc}
	\frac{(V_g+V_u) + {e_2}(V_g-V_u)}{2}& 0 \\
	0 & \frac{(V_g+V_u) - {e_2}(V_g-V_u)}{2}
	\end{array}
	\right) \;,
\label{eq:VBOJI}
\end{equation}
in which $V_{g,u}$ are the two BO potential energy curves for the $^2\Sigma_{g,u}^+$ molecular states respectively. 

For $F = 3/2$, there is only one channel that only opens when the collision energy is above the upper hyperfine threshold. The hyperfine term $V^{\textrm{hf}} = \Delta E^{\textrm{hf}}$, and the BO term is given by
\begin{equation}
V^{\textrm{BO}} = \frac{1}{2}\left[(V_g+V_u) + {e_1}(V_g-V_u)\right] \;.
\end{equation}

\begin{figure}
\includegraphics[width=\columnwidth]{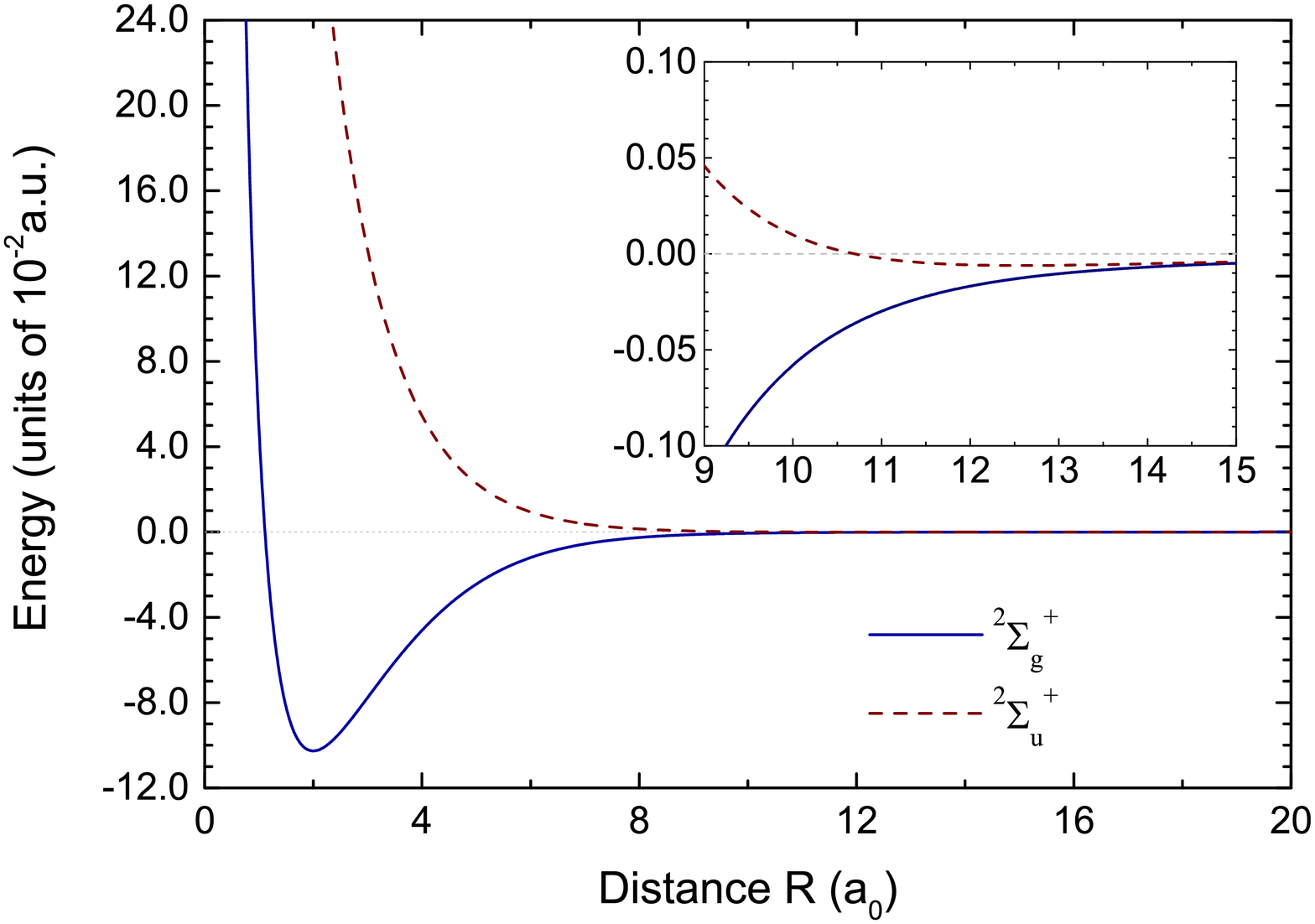}
\caption{BO potential energy curves of the \textit{gerade} (solid line) and the \textit{ungerade} (dashed line) states constructed in our work for the proton-hydrogen collision}
\label{fig:Potential}
\end{figure}

To perform the numerical calculation, we need the BO potential energy curves $V_{g,u}$. For the proton-hydrogen interaction, the potential curves can be obtained analytically in the prolate spheroidal coordinates following the method of Ref.~\cite{Hunter1966}. However, to simplify the calculation, we opt for an easier way to construct these potential curves, which is illustrated as follows. For internuclear separation $R$ from $0.4$a.u. to $10.0$a.u., we use a cubic spline~\cite{cppRec} to interpolate the data points given in Ref.~\cite{Hunter77} (we will refer to this reference as paper \textit{I}) which are calculated using the exact method of Ref.~\cite{Hunter1966}. For $R$ larger than $10.0$a.u., we use the asymptotic expansion from Ref.~\cite{Damburg1968}. More specifically,
\begin{equation}
V_{g,u} = V_0(R) \mp \frac{1}{2}\Delta V(R) \;,
\end{equation}
with $\mp$ for $^2\Sigma^+_g$ and $^2\Sigma^+_u$, respectively. Here,
\begin{equation}
V_0(R) = -\frac{9}{4R^4}-\frac{15}{2R^6}-\frac{213}{4R^7}-\frac{7755}{64R^8}-\frac{1773}{2R^9} \;,
\end{equation}
and
\begin{align}
\Delta V(R) =&\; 4R\mathrm{e}^{-R-1} \left( 1+\frac{1}{2R}-\frac{25}{8R^2}-\frac{131}{48R^3}-\frac{3923}{384R^4} \right. \nonumber\\[5pt]
&\left. -\frac{145399}{3840R^5}-\frac{521989}{46080R^6}-\frac{509102915}{645120R^7}\right. \nonumber\\[5pt]
&\left. -\frac{37749539911}{10321920R^8}\right)\;.
\end{align}
For $R$ smaller than $0.4$a.u., we use fitted functions for the inner wall, given by 
\begin{align}
V_g(R) =& \; 0.835\left(\frac{R}{R_0}+0.0012\right)^{-1.0597}-0.993 \;,\\
V_u(R) =& \; 0.932\left(\frac{R}{R_0}-0.00009\right)^{-1.031}+0.0896 \;, 
\end{align}
in which $R_0 = 1.000544628$ atomic unit. All the potential equations are in atomic units. The two potential energy curves that we constructed are shown in Fig.~\ref{fig:Potential}. Notice that the \textit{ungerade} state has a very shallow well with the depth around $0.0001$ atomic unit. Despite the shallow potential well and the light reduced mass, the \textit{ungerade} potential supports two molecular bound states~\cite{Car04,Kai13}, one of which is extremely weakly bounded with the bound state energy around $10^{-9}$a.u.~\cite{Car04,Kai13}. The extremely-near-threshold bound state of the \textit{ungerade} potential suggests there may be a resonance very near the threshold, which makes the evaluation of the $s$-wave scattering length very sensitive to the potential in the inner region~\cite{Gao2013c}. This may explain the relatively large discrepancy of the $s$-wave scattering length for the \textit{ungerade} state reported by previous studies despite the availability of an analytical potential~\cite{Car04,Glassgold05,bod08,Kai13}.

In this work, the CC equations are integrated numerically using a hybrid propagator~\cite{man86, ale87} constructed similarly to the one used in the Hibridon scattering code~\cite{hibridon}. It employs a modified version of the log-derivative method of Johnson~\cite{joh73} by Manolopoulos~\cite{man86} at short range, and a modified version of the potential-following method of Gordon~\cite{gor69} by Alexander and Manolopoulos~\cite{ale87} at long range. Convergence can be tested on the resulting $S$ matrix after being converted from the log-derivative matrix following the method of Johnson~\cite{joh73}.

\subsection{Elastic approximation}
\label{sec:elastic}

At energies that are much greater than the atomic hyperfine splitting, the hyperfine interaction can be neglected to simplify the theory. Without the hyperfine interation, the asymptotic states in the $JI$ channels are degenerate, and the potential matrix is diagonal in the $JI$ channels in all internuclear separations. Thus, instead of solving the multichannel CC equations in the $FF$ channels, we can solve the single-channel radial equations independently in the $JI$ channels to obtain the phase shifts for each channel. These phase shifts constitute the physical $K$ matrix in the $JI$ channels, $K^{Fl(JI)}$, which can then be transformed to the physical $K$ matrix in the $FF$ channels, $K^{Fl}$, and converted to the $S$ matrix. Such is the essense of the elastic approximation~\cite{dal61}, as discussed in Ref.~\cite{zyg03}.

Under the elastic approximation, for $1/2\le F \le 2I_1-1/2$, the physical $K$ matrix in the $JI$ channels can be written as
\begin{equation}
K^{Fl(JI)} =
	\left(
	\begin{array}{cc}
	\frac{(K_g+K_u) + {e_2}(K_g-K_u)}{2}& 0 \\
	0 & \frac{(K_g+K_u) - {e_2}(K_g-K_u)}{2}
	\end{array}
	\right) \;,
\label{eq:KJI}
\end{equation} 
where $K_{g,u} (\epsilon,l) = \tan\delta_l^{g,u}(\epsilon)$ with $\delta_l^{g,u}(\epsilon)$ being the single-channel phase shifts of the $l$th partial wave for the \textit{gerade} state and the \textit{ungerade} state respectively~\cite{LG2012}. These single-channel phase shifts can be calculated from solving the single-channel equations using numerical methods or single-channel QDT~\cite{LG2012}. Then the physical $K$ matrix in the $FF$ channels can be obtained from
\begin{equation}
K^{Fl} = U^{F\dagger}K^{Fl(JI)}U^{F} \;,
\label{eq:KFF}
\end{equation} 
and the $S$ matrix can be obtained from Eq.~(\ref{eq:Smatrix}). For $F=2I_1+1/2$, there is only one channel, so the approximation does not apply. 

Applying the elastic approximation using Eqs.~(\ref{eq:KJI}), (\ref{eq:KFF}), and (\ref{eq:Smatrix}), different cross sections can be expressed in terms of $K_{g,u}(\epsilon,l)$ or $\delta_l^{g,u}(\epsilon)$. Here we present the simplified expressions of the total cross sections given in Eq.~(\ref{eq:totalcs}) under the elastic approximation. For the total cross sections, the zero energy in the elastic approximation, with consideration of the nuclear spin statistics, should be offset from the zero energy in the multichannel treatment by the center of gravity, which is given by $(I_1+1)/(2I_1+1)\cdot\Delta E^{hf}$. 

The hyperfine de-excitation cross section simplifies to
\begin{align}
\sigma_{\textrm{de}} \stackrel{\epsilon \gg \Delta E^{\textrm{hf}}}{\sim} &\frac{I_1}{2I_1+1}\frac{\pi}{k^2} \sum_{l=0}^{\infty}(2l+1)\frac{(K_g-K_u)^2}{(1+K_g^2)(1+K_u^2)}\nonumber\\
=&\frac{I_1}{2I_1+1}\frac{\pi}{k^2} \sum_{l=0}^{\infty}(2l+1)\sin^2(\delta^{u}_l-\delta^{g}_l) \;,
\label{eq:ElaApprox}
\end{align}
where $\sigma_{\textrm{de}}$ is short for $\sigma(\{I_1+1/2,I_1\}\rightarrow\{I_1-1/2,I_1\})$. This is consistent with the result given in Ref.~\cite{dal65}. If we define the spin exchange cross section~\cite{dal65} (same as the charge exchange cross section defined in Ref.~\cite{cot00}) as
\begin{equation}
\sigma_{\textrm{se}} \equiv \frac{\pi}{k^2} \sum_{l=0}^{\infty}(2l+1)\sin^2(\delta^{u}_l-\delta^{g}_l) \;,
\end{equation} 
Eq.~(\ref{eq:ElaApprox}) becomes
\begin{equation}
\sigma_{\textrm{de}} \stackrel{\epsilon \gg \Delta E^{\textrm{hf}}}{\sim} \frac{I_1}{2I_1+1} \sigma_{\textrm{se}} \;.
\label{eq:EAde}
\end{equation} 
The coefficient $I_1/(2I_1 + 1)$ takes into account of the nuclear spin statistics, and equals $1/4$ for the proton-hydrogen interaction.

The corresponding hyperfine excitation cross section is related to the de-excitation cross section by a detailed balance relation guaranteed by the time-reversal symmetry~\cite{lan77}, which is given by
\begin{equation}
\frac{\sigma_{\textrm{ex}}}{\sigma_{\textrm{de}}} = \frac{I_1+1}{I_1}\cdot \frac{\epsilon-\Delta E^{\textrm{hf}}}{\epsilon} \stackrel{\epsilon \gg \Delta E^{\textrm{hf}}}{\sim} \frac{I_1+1}{I_1} \;.
\label{eq:exde}
\end{equation}
$\sigma_{\textrm{ex}}$ is short for $\sigma(\{I_1-1/2,I_1\}\rightarrow\{I_1+1/2,I_1\})$, and can be written in terms of the spin exchange cross section as
\begin{equation}
\sigma_{\textrm{ex}} \stackrel{\epsilon \gg \Delta E^{\textrm{hf}}}{\sim} \frac{I_1+1}{2I_1+1} \sigma_{\textrm{se}} \;.
\end{equation}
Therefore, for the proton-hydrogen system, the coefficient in front of $\sigma_{\textrm{se}}$ is $3/4$.

For the total elastic cross sections under the elastic approximation, we define the elastic partial wave cross sections
\begin{align}
\sigma^{g,u}_l &\equiv \frac{4\pi}{k^2}(2l+1)\frac{K_{g,u}^2}{1+K_{g,u}^2} \nonumber\\
&= \frac{4\pi}{k^2}(2l+1)\sin^2\delta^{g,u}_l \;,
\end{align}
for the \textit{gerade} state and the \textit{ungerade} state respectively. We also define the total elastic cross sections as the summation of the corresponding partial wave cross sections over $l$, given by
\begin{equation}
\sigma^{g,u}_{\textrm{tot}} \equiv \sum_{l=0}^{\infty}\sigma^{g,u}_l \;.
\end{equation}
The total elastic cross section in the lower hyperfine channel, $\sigma_{\textrm{lo}} \equiv \sigma(\{I_1-1/2,I_1\}\rightarrow\{I_1-1/2,I_1\})$, is given by
\begin{align}
\sigma_{\textrm{lo}} \stackrel{\epsilon \gg \Delta E^{\textrm{hf}}}{\sim} &\frac{1}{2}(\sigma^{g}_{\textrm{tot}}+\sigma^{u}_{\textrm{tot}}) + \frac{1}{4I_1+2}\sum_{l=0}^{\infty}e_1(\sigma^{g}_l-\sigma^{u}_l) \nonumber\\
&- \frac{I_1+1}{2I_1+1} \sigma_{\textrm{se}}\;.
\end{align}
And the total elastic cross section in the higher hyperfine channel, $\sigma_{\textrm{hi}} \equiv \sigma(\{I_1+1/2,I_1\}\rightarrow\{I_1+1/2,I_1\})$, is given by
\begin{align}
\sigma_{\textrm{hi}} \stackrel{\epsilon \gg \Delta E^{\textrm{hf}}}{\sim} &\frac{1}{2}(\sigma^{g}_{\textrm{tot}}+\sigma^{u}_{\textrm{tot}}) + \frac{1}{4I_1+2}\sum_{l=0}^{\infty}e_1(\sigma^{g}_l-\sigma^{u}_l) \nonumber\\
&- \frac{I_1}{2I_1+1} \sigma_{\textrm{se}}\;.
\end{align}


\section{Results and discussions}
\label{sec:Results}


\subsection{Multichannel treatment vs. elastic approximation}
\label{sec:MvE}

\begin{figure}
\includegraphics[width=\columnwidth]{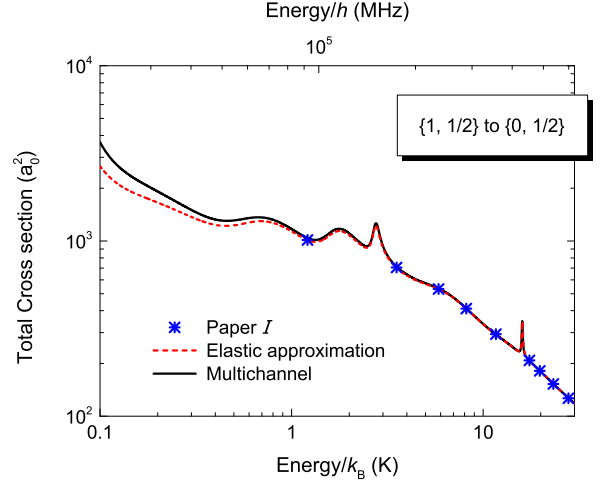}
\caption{Comparison of the spin-exchange cross section from paper \textit{I} multiplied by $1/4$ that accounts for the nuclear statistics (stars) and the total hyperfine de-excitation cross section from channel $\{ F_1 = 1, F_2 = 1/2 \}$ to channel $\{ F_1 = 0, F_2 = 1/2 \}$ using the multichannel treatment (solid line) and the elastic approximation (dashed line).
\label{fig:10TotalCS-Elastic}}
\end{figure}

We start with comparing our elastic approximation results of the hyperfine de-excitation cross section with the spin exchange cross sections given in paper \textit{I}, as well as with results from our multichannel treatment. According to Eq.~\ref{eq:EAde}, the spin exchange cross section is four times the hyperfine de-excitation cross section under the elastic approximation for the proton-hydrogen interaction. Therefore, we compare our de-excitation cross section results with the spin exchange cross sections given in paper \textit{I} multiplied by $1/4$. Also, all the cross section results from the elastic approximation need to be offset in energy by the center-of-gravity $(I_1+1)/(2I_1+1)\cdot\Delta E^{hf} = 0.0511265$K. Our results from both the elastic approximation and the multichannel treatment are obtained using the numerical calculation.

Fig.~\ref{fig:10TotalCS-Elastic} shows the hyperfine de-excitation cross sections, where the data points converted from the results given in paper \textit{I} lie right on top of our elastic approximation results. This demonstrates the validity of our potential energy curves and numerical method. The results from the multichannel treatment and the elastic approximation agree almost exactly for energies between $5$ and $30$ kelvin. The discrepancy between the two becomes more significant when energy decreases especially below $1$ kelvin. 

\begin{figure}
\includegraphics[width=\columnwidth]{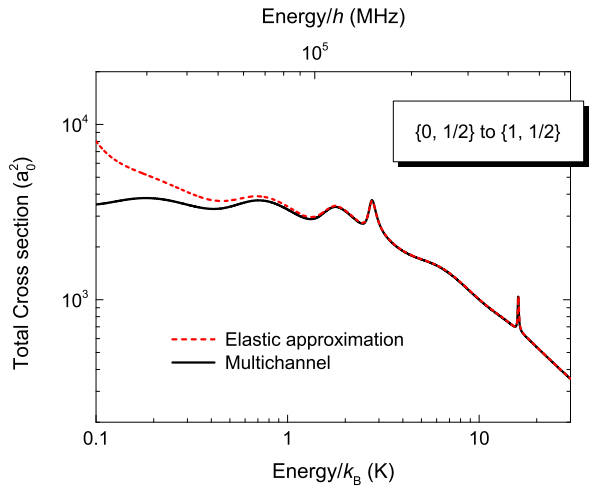}
\caption{Total hyperfine excitation cross sections from channel $\{ F_1 = 0, F_2 = 1/2 \}$ to channel $\{ F_1 = 1, F_2 = 1/2 \}$ from the elastic approximation (dashed line) and the multichannel treatment (solid line).
\label{fig:01TotalCS-Elastic}}
\end{figure}

\begin{figure}
\includegraphics[width=\columnwidth]{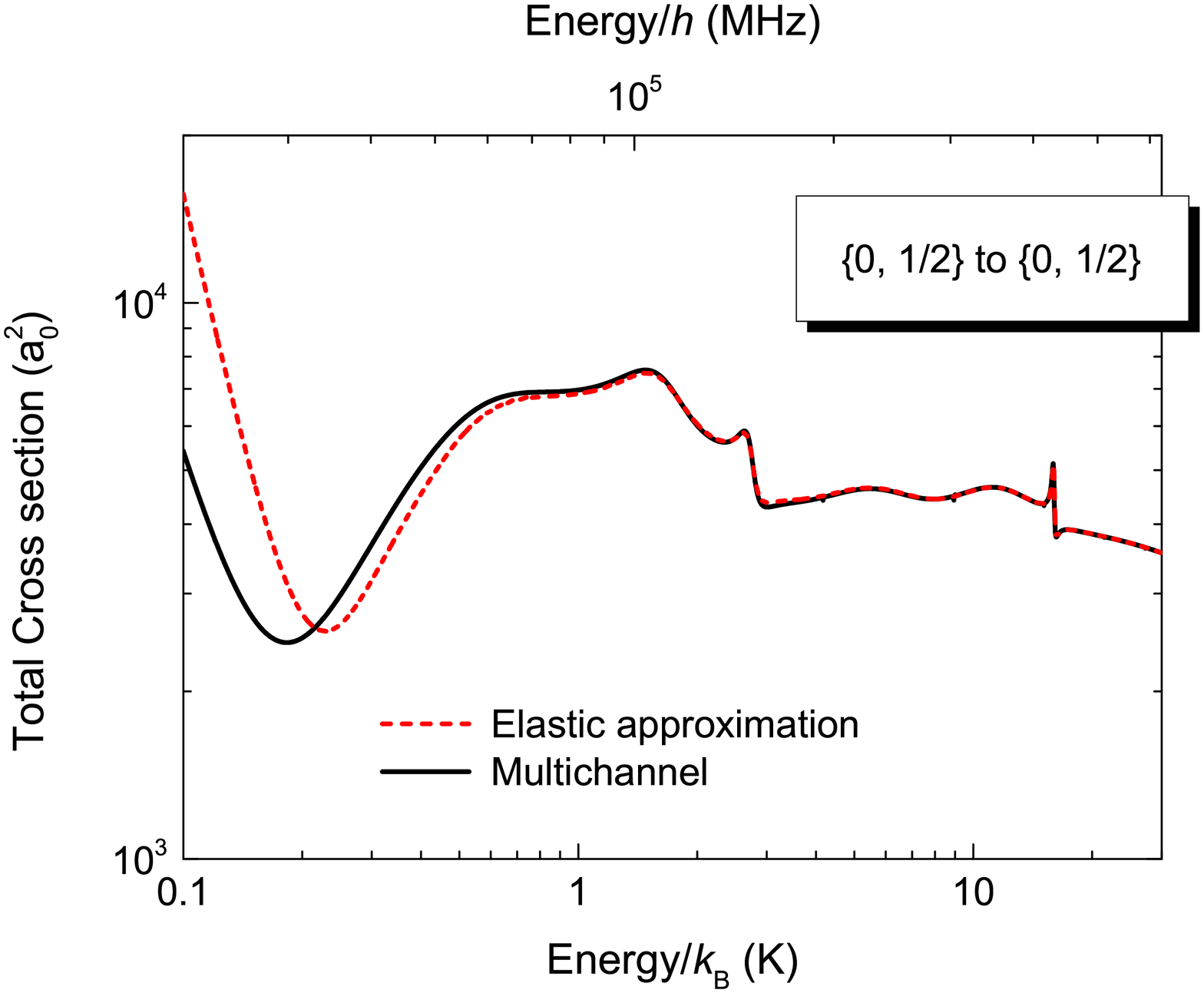}
\caption{Total elastic cross sections in the lower hyperfine channel $\{ F_1 = 0, F_2 = 1/2 \}$ from the elastic approximation (dashed line) and the multichannel treatment (solid line).
\label{fig:00TotalCS-Elastic}}
\end{figure}

\begin{figure}
\includegraphics[width=\columnwidth]{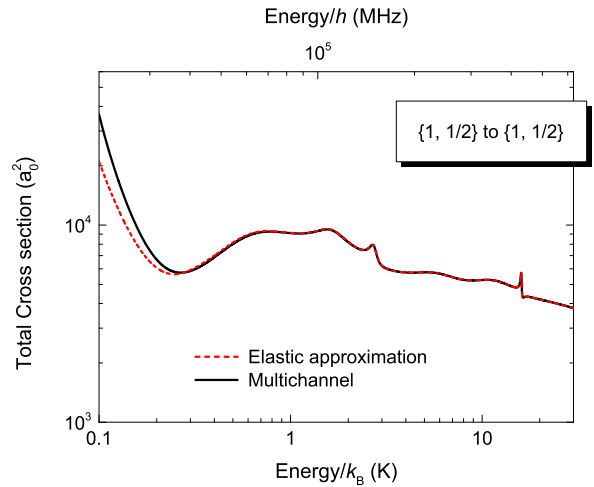}
\caption{Total elastic cross sections in the upper hyperfine channel $\{ F_1 = 1, F_2 = 1/2 \}$ from the elastic approximation (dashed line) and the multichannel treatment (solid line).
\label{fig:11TotalCS-Elastic}}
\end{figure}

Figures~\ref{fig:01TotalCS-Elastic}, \ref{fig:00TotalCS-Elastic}, and \ref{fig:11TotalCS-Elastic} shows the hyperfine excitation cross sections, the elastic cross sections in the lower hyperfine channel, and the elastic cross sections in the higher hyperfine channel respectively. These comparisons, along with the hyperfine de-excitation case, show that, for the proton-hydrogen interaction, the elastic approximation is applicable for energies larger than $1$K, and becomes more accurate with higher energy. However, it fails when hyperfine interaction is comparable or smaller than the total energy. Thus for temperatures from ultracold up to about $1$K, which is approximately an order of magnitude larger than the hyperfine splitting, the elastic approximation is not applicable, and multichannel treament with proper accounting of the hyperfine interaction should be used.

\begin{figure}	
\includegraphics[width=\columnwidth]{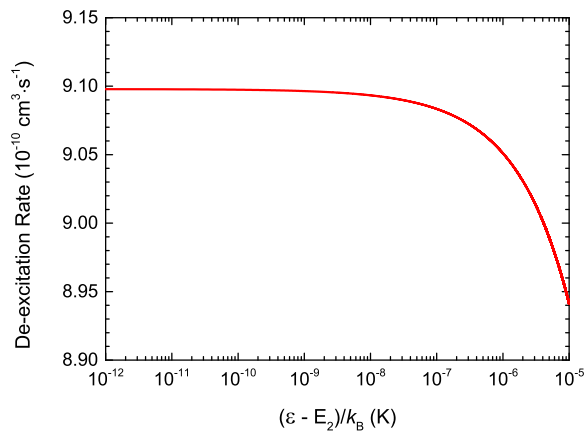}
\caption{Threshold behavior of the hyperfine de-excitation rate $\mathcal{W}_{de}$ just above the upper threshold $E_2$. The $x$-axis represents the temperature equivalence of the initial kinetic energy $(\epsilon - E_2)/k_{\mathrm{B}}$. The results are produced using the multichannel numerical calculation.
\label{fig:10Rate}}
\end{figure}

The elastic approximation also gives the incorrect threshold behavior at the upper hyperfine threshold. In previous attempts to extend elastic approximation to lower energy with effective range theory, a constant spin exchange cross-section with the corresponding rate approaching zero was predicted when energy decreases~\cite{Glassgold05, Furlanetto2007, bod08}. With multichannel treatment that incorporates the hyperfine splitting, the de-excitation cross section follows the Wigner's threshold law~\cite{wig48} which diverges as $(\epsilon-E_2)^{-1/2}$ above the upper hyperfine threshold. Therefore, the hyperfine de-excitation rate without thermal averaging,
\begin{equation}
\mathcal{W}_{\mathrm{de}} \equiv v\sigma_{\mathrm{de}} = \sqrt{\frac{2(\epsilon-E_2)}{\mu}}\sigma_{\mathrm{de}} \;,
\end{equation}
will reach a constant when the relative velocity $v$ approaches zero. As illustrated in Fig.~\ref{fig:10Rate} with results from multichannel numerical calculation, the de-excitation rate rises while initial kinetic energy decreases, until eventually reaches a plateau with a constant rate of $9.098\times 10^{-10} \mathrm{cm}^3\cdot\mathrm{s}^{-1}$ approximately. The same threshold behavior should also be present for electron-hydrogen and hydrogen-hydrogen collisions.

\subsection{MQDT vs. Numerical calculation}
\label{sec:NvM}

\begin{figure}	
\includegraphics[width=\columnwidth]{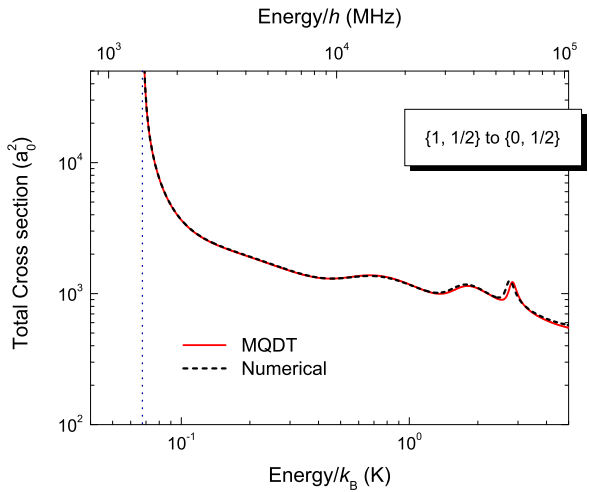}
\caption{Total hyperfine de-excitation cross sections from channel $\{ F_1 = 1, F_2 = 1/2 \}$ to channel $\{ F_1 = 0, F_2 = 1/2 \}$ from MQDT (solid line) and numerical method (dashed line). The vertical dashed line identifies the upper hyperfine threshold located at $\epsilon_2/k_B \approx 0.0682$ K, around which the cross section diverges as $(\epsilon-E_2)^{-1/2}$. 
\label{fig:10TotalCS}}
\end{figure}

In this subsection, we present the total cross section results from ultracold temperature to $5$K using multichannel treatment. In addition, we compare the results from MQDT with the numerical calculation to demonstrate the applicability of MQDT in this temperature regime.

Figure~\ref{fig:10TotalCS} shows the total cross sections for hyperfine de-excitation process from the upper hyperfine threshold to $5$ kelvin. Notice that, as mentioned in the previous subsection, the de-excitation cross section follows the Wigner's threshold law~\cite{wig48}, which diverges as $(\epsilon-E_2)^{-1/2}$ above the upper hyperfine threshold. MQDT results are almost exactly on top of the numerical calculation below one kelvin, and, although there is slight deviation, the two methods still agrees relatively well from one to five kelvin. The discrepancy is in general within one percent which can be attributed to slight energy dependence of the short-range QDT parameters. 

\begin{figure}
\includegraphics[width=\columnwidth]{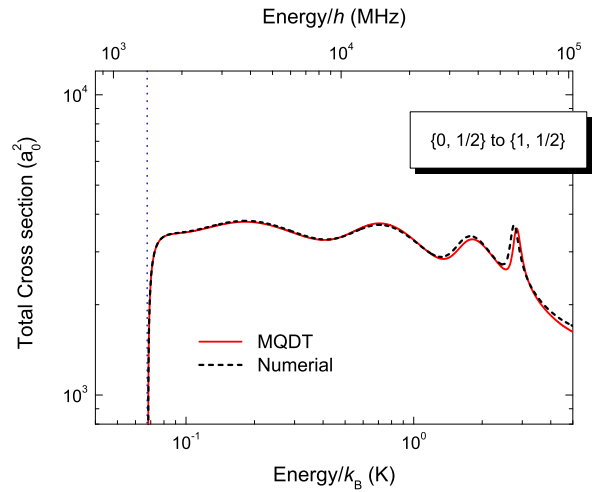}
\caption{ Total hyperfine excitation cross sections from channel $\{ F_1 = 0, F_2 = 1/2 \}$ to channel $\{ F_1 = 1, F_2 = 1/2 \}$ from MQDT (solid line) and numerical method (dashed line). The vertical dashed line identifies the upper hyperfine threshold located at $\epsilon_2/k_B \approx 0.0682$ K, around which the cross section behaves as $(\epsilon-E_2)^{1/2}$. 
\label{fig:01TotalCS}}
\end{figure}

Figure~\ref{fig:01TotalCS} shows the total cross sections for hyperfine excitation in which the Hydrogen atom is excited from its $F_1=0$ hyperfine state to its $F_1=1$ hyperfine state. It behaves as $(\epsilon-E_2)^{1/2}$ above the upper threshold. The excitation cross section is related to the de-excitation cross section by Eq.~(\ref{eq:exde}) which is guaranteed by the time-reversal symmetry~\cite{lan77}. 

\begin{figure}
\includegraphics[width=\columnwidth]{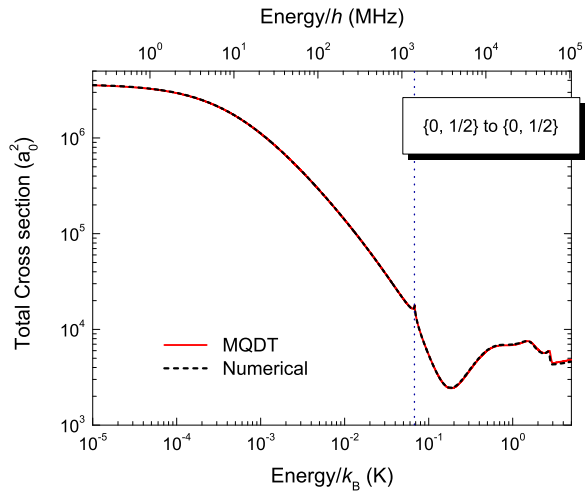}
\caption{ Total elastic cross sections in the lower channel $\{ F_1 = 0, F_2 = 1/2 \}$ from MQDT (solid line) and numerical method (dashed line). The vertical dashed line identifies the upper hyperfine threshold at $0.0682$ K. 
\label{fig:00TotalCS}}
\end{figure}

\begin{figure}
\includegraphics[width=\columnwidth]{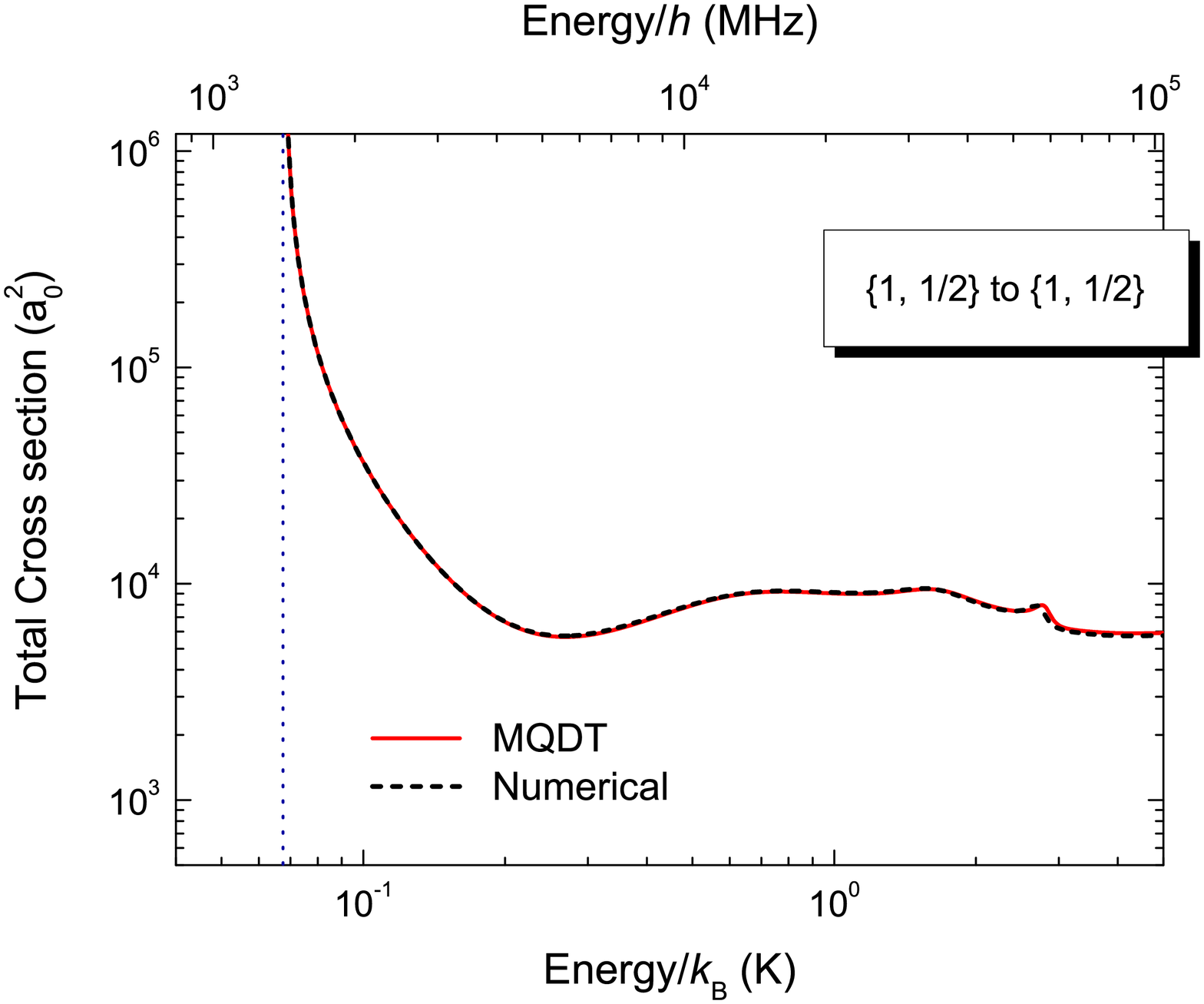}
\caption{ Total elastic cross sections in the upper channel $\{ F_1 = 1, F_2 = 1/2 \}$ from MQDT (solid line) and numerical method (dashed line). The vertical dashed line identifies the upper hyperfine threshold at $0.0682$ K. 
\label{fig:11TotalCS}}
\end{figure}

Figure~\ref{fig:00TotalCS} and ~\ref{fig:11TotalCS} depict the total cross sections for elastic scattering in the lower hyperfine channel $\{ F_1 = 1, F_2 = 3/2 \}$ and the higher hyperfine channel $\{ F_1 = 1, F_2 = 3/2 \}$ respectively, in which the atom remains in the same hyperfine level after the scattering while its $M_1$ may or may not change. The elastic cross-section in the higher hyperfine channel, just like the de-excitation cross section, diverges as $(E-E_2)^{-1/2}$ above the upper threshold, implying a constant rate in the zero temperature limit. 

Compared to our previous results on the resonant charge exchange problem of $^{23}\mathrm{Na} + ^{23}\mathrm{Na}^+$~\cite{LYG2014}, there are not any significant resonance structure within the hyperfine splitting for the proton-hydrogen system, even though the hyperfine splittings of the two systems are of the same order of magnitude. This can be attributed to the small atomic mass and polarizability of hydrogen which give an unusually large energy scale, $s_E = 0.0416$K for the proton-hydrogen system compared to other ion-atom systems ($s_E = 2.21$ $\mu\mathrm{K}$ for the sodium system). With the large energy scale, much fewer partial waves contribute within the hyperfine splitting with much smaller possibility of encountering a resonance.

Figures~\ref{fig:01TotalCS}, \ref{fig:00TotalCS}, and \ref{fig:11TotalCS} are similar to Fig.~\ref{fig:10TotalCS} in showing the agreement between results from MQDT and the numerical calculation, and the conclusion we drew for the de-excitation cross section case also stands for the others. This again demonstrates the capability of MQDT to accurately characterize multichannel ion-atom interactions from zero energy up to several kelvin with only a handful of parameters (five in this case), as we have demonstrated before on the example of sodium resonant charge exchange~\cite{LYG2014}. Also examined and verified by these comparisons are the physical picture behind our MQDT formulation, that in this energy regime, the energy and partial wave dependences are primarily due to the long-range interaction which can be accurately characterized by the analytic solution of the long-range potentials, and the short-range interaction is energy and partial wave insensitive~\cite{gao05a, Gao2013c}. 

Computationally, MQDT is much more efficient than the numerical calculation even when the QDT functions are calculated on the fly. Since the QDT functions are universal mathematical functions that are the same for all applications and can be computed to arbitrary precision with efficient algorithms \cite{Gao2013c}, their computation can be further accelerated to be as efficient as most other mathematical special functions. This computational advantage will be even more pronounced in other applications with more coupled channels.

\section{Summary}
\label{sec:Summary}

We have presented the calculation of the total cross sections of the proton-hydrogen interaction in the cold and ultracold regime. First, we demonstrated the breaking down of the elastic approximation by the comparisons with the multichannel treatment. The elastic approximation results start to deviate from multichannel treatment results when temperature decreases, especially below one kelvin, which is approximately an order of magnitude larger than the hyperfine splitting. This confirms that the hyperfine interaction becomes important for total energies comparable or smaller than the atomic hyperfine splitting, and the proper multichannel treatment has to be applied. The multichannel nature of the hyperfine structure also leads to a different threshold behavior compared to that predicted by the elastic approximation above the upper hyperfine threshold. We then present the new total cross section results using MQDT with the five-parameter implementation from ultracold temperature to five kelvin. The traditional multichannel numerical calculation confirms the MQDT results throughout the whole energy range with discrepancies smaller than one percent on average. This demonstrates again, in addition to our previous work~\cite{LYG2014}, that MQDT gives a complete understanding and characterization of the physics of ion-atom interaction, especially in cold and ultracold temperatures where quantum effects are important, which can be the key to the systematic understanding of quantum few-body systems, chemical reactions, and many-body systems involving ions.

\begin{acknowledgments}
We thank Li You and Meng Khoon Tey for helpful discussions. This work was supported in part by NSF under Grant No. PHY-1306407. BG also acknowledges partial support by  NSFC under Grant No.~11328404.
\end{acknowledgments}

\bibliography{bgao,qdt,ionAtom,twobody,fewbody,manybody,atom,numerical,astrochem,mli}

\end{document}